\documentclass[preprint,prd,superscriptaddress]{revtex4}

\usepackage{graphicx}
\usepackage{dcolumn}
\usepackage{bm}

\begin{document}

\preprint{KEK-CP-178}
\preprint{RIKEN-TH-74}
\preprint{YITP-06-34}

\title{
  Lattice gauge action suppressing near-zero modes of $H_W$
}

\newcommand{\RIKEN}{
  Theoretical Physics Laboratory, RIKEN,
  Wako 351-0198, Japan
}

\newcommand{\YITP}{
  Yukawa Institute for Theoretical Physics, 
  Kyoto University, Kyoto 606-8502, Japan
}

\newcommand{\HUDP}{
  Department of Physics, Hiroshima University,
  Higashi-Hiroshima 739-8526, Japan
}

\newcommand{\GUAS}{
  School of High Energy Accelerator Science,
  The Graduate University for Advanced Studies (Sokendai),
  Tsukuba 305-0801, Japan
}

\newcommand{\KEK}{
  High Energy Accelerator Research Organization (KEK),
  Tsukuba 305-0801, Japan
}

\author{Hidenori~Fukaya}
\affiliation{\RIKEN}

\author{Shoji~Hashimoto}
\affiliation{\KEK}
\affiliation{\GUAS}

\author{Ken-Ichi~Ishikawa}
\affiliation{\HUDP}

\author{Takashi~Kaneko}
\affiliation{\KEK}
\affiliation{\GUAS}

\author{Hideo~Matsufuru}
\affiliation{\KEK}

\author{Tetsuya~Onogi}
\affiliation{\YITP}

\author{Norikazu~Yamada}
\affiliation{\KEK}
\affiliation{\GUAS}

\collaboration{JLQCD collaboration}
\noaffiliation


\begin{abstract}
  We propose a lattice action including unphysical
  Wilson fermions with a negative mass $m_0$ of the
  order of the inverse lattice spacing.
  With this action, the exact zero mode of the
  hermitian Wilson-Dirac operator $H_W(m_0)$ cannot appear
  and near-zero modes are strongly suppressed.
  By measuring the spectral density $\rho(\lambda_W)$, we
  find a gap near $\lambda_W=0$ on the configurations
  generated with the standard and improved gauge actions.
  This gap provides a necessary condition for the proof of
  the exponential locality of the overlap-Dirac operator by
  Hernandez, Jansen, and L\"uscher.
  Since the number of near-zero modes is small, the
  numerical cost to calculate the matrix sign function of
  $H_W(m_0)$ is significantly reduced, and the simulation
  including dynamical overlap fermions becomes feasible.
  We also introduce a pair of twisted mass pseudo-fermions
  to cancel the unwanted higher mode effects of the Wilson
  fermions. 
  The gauge coupling renormalization due to the additional
  fields is then minimized.
  The topological charge measured through the index of the
  overlap-Dirac operator is conserved during continuous
  evolutions of gauge field variables.
\end{abstract}

\maketitle

\section{Introduction}
In the construction of the lattice chiral fermions, the
conventional Wilson-Dirac operator $D_W$ still plays a
crucial role. 
In the domain-wall fermion
\cite{Kaplan:1992bt,Shamir:1993zy,Furman:1994ky}
the lattice Dirac operator is nothing but the Wilson-Dirac
operator in four dimensions, but the fermion field
has interactions also with its fifth dimensional neighbors. 
The overlap-Dirac operator \cite{Neuberger:1998wv}
$D$ contains a matrix sign function of the
hermitian Wilson-Dirac operator $H_W=\gamma_5 D_W$ as
\begin{equation}
  \label{eq:ov}
  D = \frac{1}{\bar{a}}
  [ 1 + \gamma_5 \mathrm{sgn}(a H_W) ], \;\;\;\;
  \bar{a} \equiv \frac{a}{1+s}.
\end{equation}
An important difference from the usual Wilson fermion is
that the mass term is given with a negative value
$m_0=-(1+s)/a$ of the order of the inverse lattice spacing
$1/a$. 
In the limit of vanishing gauge coupling, the parameter $s$
must be between $-1$ and 1 in order to obtain the single
flavor massless fermion; the sign function is well defined
since there is a lower bound in the eigenvalue spectrum of
$|H_W|$. 
In the presence of the gauge interaction, $H_W$ could
develop zero or near-zero eigenvalues, which makes the the
matrix sign function singular.
The near-zero mode does not exist for sufficiently smooth
background gauge fields $\{U_\mu(x)\}$ satisfying a condition 
$||1-P_{\mu\nu}(x)||<\epsilon$ for all plaquette variables
$P_{\mu\nu}(x)$ with $\epsilon$ a small number less than
$\sim 1/20.49$ \cite{Neuberger:1999pz}.
In the actual numerical simulations, however,
$||1-P_{\mu\nu}(x)||$ is larger by an order of magnitude,
and the near-zero modes appear quite frequently.

The origin of the near-zero modes is understood as a
local lump of the background gauge field (or so-called the
dislocation).
An analytic example of such a gauge configuration and its
associated exact zero mode is given in \cite{Berruto:2000fx}.
Because such a zero mode is localized in space-time, the
number of the near-zero modes increases as the lattice
volume $V$ is increased. 
In other words, the spectral density $\rho(\lambda_W)$ of
$H_W$ is non-zero at $\lambda_W=0$, which is true at any
finite value of gauge coupling \cite{Edwards:1998sh}.
The localization property of the near-zero modes has
recently been studied extensively 
\cite{Golterman:2003qe,Golterman:2004cy,Golterman:2005fe},
and it is found that they are exponentially localized unless 
one enters the Aoki phase, where the flavor-parity symmetry
is spontaneously broken \cite{Aoki:1983qi}.
Since the radius of the exponential fall-off is of the
order of lattice spacing $a$, the effect of the near-zero
modes disappears in the continuum limit, and therefore is a
lattice artifact.

The effect of the near-zero modes appears as a small
residual breaking of chiral symmetry in the domain-wall
formulation 
\footnote{
  For the domain-wall fermion the relevant operator
  is not $H_W$ itself but a logarithm of the transfer matrix
  in the fifth dimension.
  They share the zero modes, and the near-zero modes of the
  both operators are closely related with each other.
}.
Namely, the four-dimensional fermion mode receives
additive mass renormalization $m_{res}$ when the lattice
extent in the fifth dimension $L_5$ is finite
\cite{Aoki:2000pc,Blum:2000kn}.
The problem is not just that $m_{res}$ is finite, but the
suppression is only by $1/L_5$ rather than by $\exp(-cL_5)$
as expected for the extended non-zero modes
\cite{Golterman:2005fe,Christ:2005xh}.
For the overlap fermion, the residual mass can be made
arbitrarily small by projecting out the near-zero mode and
treating them exactly when one calculates the matrix sign
function.
The problem however manifests itself in the locality of the
overlap-Dirac operator.
The locality is proved only when there exists a gap in the
spectral density of $H_W$ near zero \cite{Hernandez:1998et}.
Therefore, the existence of the near-zero mode persisting in
the infinite volume limit could potentially spoil the
locality of the overlap-Dirac operator and thus the
controlled continuum limit of the overlap fermion.
Even if there exists non-zero density of the near-zero
modes, the locality may still be maintained
\cite{Golterman:2003qe}, provided that the near-zero modes
are exponentially localized and its exponential fall-off is
sufficiently fast.
The effect of the near-zero modes becomes irrelevant in the
continuum limit, as the localization length scales as the
lattice spacing $a$.
In this case the locality of the near-zero modes should
always be checked in order to make sure that the Aoki phase
is not entered.

In this paper we propose the use of the lattice action with
two flavors of extra Wilson fermions with the large negative
mass $m_0$. 
Since the occurrence of the near-zero modes is suppressed by
the fermion determinant, we expect that the spectral
density $\rho(\lambda_W)$, as defined in
(\ref{eq:spectral_density}), vanishes at $\lambda_W=0$ and
the near-zero modes are strongly suppressed.
By suppressing the near-zero modes, the problem of locality
of the overlap-Dirac operator is essentially solved.
Furthermore, the numerical cost for applying the
overlap-Dirac operator is significantly reduced, because the
cost for projecting out the low-lying eigenmodes is
proportional to the eigenvalue density of $H_W$.
This is especially important when one wants to include the
dynamical effect of the overlap fermions in the Monte Carlo
simulations.

The idea of adding the extra Wilson fermions is very simple
and in fact has been around for several years
\cite{Vranas:1999rz,Izubuchi:2002pq,Luscher_pc}, 
but detailed numerical study has been missing until
recently.
A preliminary report of this work was included in
\cite{Fukaya:2006ca}, and a paper by Vranas
\cite{Vranas:2006zk} was submitted very recently, 
after this work had been essentially completed. 

With the action that forbids the occurrence of the zero mode,
the global topological charge of the gauge field
configuration cannot change through continuous deformation
of the gauge variable.
This is because the spectral flow of $H_W(m)$ can never
cross zero at $m=m_0$ under the continuous deformation.
Here we use the index of the overlap-Dirac operator
constructed from $H_W(m_0)$ as a definition of the
topological charge.
Unlike the measurement of the $F\tilde{F}$ operator after
some cooling, this always gives an unique integer for the
topological charge.
The property of conserving topological charge resembles
that of the continuum theory.
In the continuum theory, there is an infinitely high
barrier in the action between different topological sectors, 
but the barrier is lowered by the lattice regularization of
gauge field.
With the extra Wilson fermions the infinite potential
barrier is recovered at any finite lattice spacing, and thus
the continuum gauge field is better approximated.

For the Monte Carlo simulations including the fermion
determinant, the Hybrid Monte Carlo (HMC)
\cite{Duane:1987de} provides the most efficient algorithm. 
HMC is based on the molecular dynamics evolution of gauge
field variables under the Hamiltonian including a
pseudo-fermion action $\chi^\dagger H_W^{-2}\chi$.
The gauge field evolves with small time steps, that
approximate the continuous evolution.
When a near-zero mode of $H_W$ appears and further approaches the
zero point, the pseudo-fermion action increases rapidly and
generates a repulsive force.
This means that the topological charge cannot change during
the HMC simulation, as far as the step size is taken small
enough.
If the step size is so large that the transmission through
the potential barrier is allowed, the conservation of the 
Hamiltonian becomes poor and the acceptance rate of the
Monte Carlo would become very low.

Since the topological charge does not change, the HMC
simulation is confined in an initially given topological
sector, and the correct sampling of the $\theta=0$ (or any
finite value of $\theta$) is not possible.
This is an advantage for the simulations in the
$\epsilon$-regime of the chiral perturbation theory (ChPT)
\cite{Gasser:1987ah}, for which one needs 
the gauge ensemble in a given topological sector.
For this purpose, some of the present authors tested a
modified plaquette gauge action to suppress the topology
change \cite{Fukaya:2005cw} (see also
\cite{Bietenholz:2005rd}). 
The proposal of the present work is more solid, as it
strictly prohibits the topology change.
Out of the $\epsilon$-regime, on the other hand, the fixed
topology is a disadvantage.
But the possible errors due to the incorrect sampling of the
$\theta=0$ vacuum disappear quickly as the physical volume is
increased.

Although we aim at performing dynamical fermion simulations
using the overlap formalism, this paper focuses on the
quenched study.
Namely, the dynamical overlap fermion is switched off, while
the extra Wilson fermions are treated dynamically.
We then numerically study the spectral density of $H_W$ for
various choices of gauge actions.

The rest of this paper is organized as follows.
In Section~\ref{sec:action} we introduce the lattice actions
that we studied in this work.
An implementation in the HMC simulation is also presented.
Our numerical simulations are explained in
Section~\ref{sec:HMC_simulations}.
Section~\ref{sec:spectral_density} contains the main results
of this work, {\it i.e.} the spectral density of $H_W$ with
and without the extra Wilson fermions.
The topology conservation is a key property of the extra
Wilson fermions.
We discuss how it works
(Section~\ref{sec:topology_conservation}) and 
what is its effect on physical quantities
(Section~\ref{sec:topology_dependence}). 
In Section~\ref{sec:beta_shift}, we calculate how much
the gauge coupling is renormalized by the extra Wilson
fermions, and show that the finite renormalization can be
made small by further adding pseudo-fermions with a twisted
mass term.
Section~\ref{sec:conclusions} contains our conclusion as
well as some discussions about possible artifacts due to the
fixed topology.

\section{Lattice action and its implementation}
\label{sec:action}

We consider a lattice action
\begin{equation}
  \label{eq:action}
  S = S_G + S_E,
\end{equation}
where $S_G$ is any gauge action, such as the Wilson,
L\"uscher-Weisz, Iwasaki, {\it etc.}, 
and $S_E$ denotes the extra Wilson fermion.
In the following numerical analysis, we use for the gauge
part $S_G$ the plaquette gauge action $S_{Pl}$, with and
without a modification to suppress dislocations,
and the Renormalization Group (RG)
improved (or Iwasaki) action $S_{RG}$.
The plaquette action is written as
\begin{equation}
  \label{eq:admiaction}
  S_{Pl} = \left\{
    \begin{array}{ll}\displaystyle
      \beta\sum_{x,\mu<\nu}
      \frac{1-\mathrm{Re}\mathrm{Tr}P_{\mu\nu}(x)/3}
      {1-(1-\mathrm{Re}\mathrm{Tr}P_{\mu\nu}(x)/3)/\epsilon},
      & \;\;\; \mbox{when} \;\;
      1-\mbox{ReTr}P_{\mu\nu}(x)/3 < \epsilon,
      \\
      \infty & \;\;\; \mbox{otherwise}
    \end{array}
    \right.,
\end{equation}
where $P_{\mu\nu}(x)$ is the plaquette variable at $x$ on
the $\mu$-$\nu$ plane.
The denominator is introduced to suppress the local lump of
the gauge field for which 
$1-\mathrm{Re}\mathrm{Tr}P_{\mu\nu}(x)/3$ takes a large
value \cite{Luscher:1998du,Fukaya:2005cw,Bietenholz:2005rd}.
When $1/\epsilon=0$, it reduces to the standard Wilson gauge
action.
For this gauge action with a finite $1/\epsilon$,
$1-\mathrm{Re}\mathrm{Tr}P_{\mu\nu}(x)/3$ cannot become
larger than the parameter $\epsilon$.
If $\epsilon$ is chosen smaller than $1/20.49$, then the
hermitian Wilson-Dirac operator is proved to have a gap
\cite{Neuberger:1999pz}, but we take a larger number 3/2 in
the numerical simulations, because otherwise the lattice
spacing becomes too small even at the strong coupling limit.
Another choice of the gauge action is that of Iwasaki
\cite{Iwasaki:1983ck}, which includes the rectangular term
with the parameter $c_1=-0.331$.
The denominator as in (\ref{eq:admiaction}) is not
introduced for this case, but it is known that the
rectangular term has an effect to suppress the
dislocations. 

The extra fermion term $S_E$ is written as
\begin{equation}
  \label{eq:fermion+ghost}
  S_E = 
  \sum_x \bar{\psi}(x) D_W(m_0) \psi(x) +
  \sum_x \phi^\dagger(x) [D_W(m_0) + i\mu\gamma_5\tau_3] \phi(x),
\end{equation}
where $\psi$ denotes two flavors of extra heavy Wilson
fermion with a negative mass $m_0$.
The second term is a pseudo-fermion term introduced in order to
cancel unwanted effects of the Wilson fermion especially in
the ultraviolet region, which leads to a large shift of the
$\beta$ value to be used in the simulation as discussed
later in Section~\ref{sec:beta_shift}.
Because of an additional mass term 
$\phi^\dagger i\mu\gamma_5\tau_3\phi$,
which is twisted in the flavor space by $\tau_3$,
the extra Wilson fermion works to suppress the near-zero
modes of $H_W(m_0)$ as expected.
In fact, the action $S_E$ generates the suppression
factor 
\begin{equation}
  \label{eq:det}
  \det\left[
    \frac{H_W(m_0)^2}{H_W(m_0)^2 + \mu^2}
  \right]
\end{equation}
in the partition function.
The twisted mass $\mu$ controls the range of the near-zero
eigenvalues suppressed by the numerator.
The eigenvalues whose absolute value is lower
than $\mu$ are strongly suppressed, while the other higher
modes are less affected.
The limit of $\mu\to\infty$ corresponds to the case where
the pseudo-fermion term is switched off.
When $\mu=0$, the cancellation is exact, and there is no
extra fermions and pseudo-fermions.

Since the action (\ref{eq:action}) includes the fermion,
some dynamical fermion algorithm is needed to generate the
gauge field ensembles.
Application of the HMC algorithm is straightforward except
for the additional boson term.
In order to cancel the higher modes of $H_W$ efficiently, we
use only one pseudo-fermion for both fields.
Namely, the Hamiltonian for the molecular dynamics evolution
contains a term
\begin{equation}
  \label{eq:pseudo-fermion}
  \sum_x \chi^\dagger(x)
  \left\{
    [D_W(m_0)+i\mu\gamma_5] 
    [D_W(m_0)]^{-1} [D_W^\dagger(m_0)]^{-1}
    [D_W^\dagger(m_0)-i\mu\gamma_5]
  \right\} \chi(x)
\end{equation}
with the pseudo-fermion field $\chi$.
Then, the fermion force derived from
(\ref{eq:pseudo-fermion}) largely cancels in the
combination $[D_W(m_0)+i\mu\gamma_5][D_W(m_0)]^{-1}$, when
the twisted mass $\mu$ is small.

In the molecular dynamics evolution with
(\ref{eq:pseudo-fermion}), an inversion of the Wilson-fermion
matrix $D_W^\dagger(m_0)D_W(m_0)$ is necessary, and
therefore it costs much more than the usual gauge action.
The cost is, however, not substantial compared with the
inversion of the overlap-Dirac operator, for instance,
needed for the dynamical overlap fermion simulation.
The cost for the inversion of $H_W^2$ is proportional to the
inverse of lowest-lying eigenvalue $\lambda_{min}^2$,
which is lifted by the introduction of the suppression
factor (\ref{eq:det}).
It means that the cost does not increase arbitrarily, even
though there is no explicit lower limit on the lowest-lying
eigenvalue. 

\section{HMC simulations}
\label{sec:HMC_simulations}
We performed Monte Carlo simulations including the extra
Wilson fermions.
Although the fermions are included, they are unphysical and
irrelevant in the continuum limit.
The physical sea quarks are not included.

The numerical simulations have been done on a 
$16^3\times 32$ lattice with three choices of the gauge
actions.
\begin{enumerate}
\item $S_{Pl}$ with $1/\epsilon=0$, the standard Wilson
  gauge action.
\item $S_{Pl}$ with $1/\epsilon=2/3$.
  With this choice the plaquette variable $P_{\mu\nu}(x)$ 
  can take any value in the SU(3) gauge group except for the
  two points $e^{\pm i 2\pi/3} I$ with $I$ the $3\times 3$
  unit matrix.
  At these isolated two points,
  $\mathrm{Re}\mathrm{Tr}P_{\mu\nu}(x)/3$ becomes minimum.
  The positivity violation as argued in \cite{Creutz:2004ir}
  occurs for $\epsilon$ smaller than 3/2.
\item $S_{RG}$, the Iwasaki gauge action.
\end{enumerate}
The simulation parameters are listed in
Table~\ref{tab:param}.
For each of the three choices of the gauge action we take three
values of $\mu$, the twisted mass of the extra
pseudo-fermions, to be 0, 0.2, and 0.4.
The large negative mass $m_0$ is always set to $-1.6$.
The gauge coupling $\beta$ is chosen such that the lattice
spacing determined through the Sommer scale $r_0$ 
($\simeq$ 0.49~fm from phenomenological models) is roughly
tuned to 0.125~fm, which corresponds to $r_0/a=3.9$.

All the simulations are carried out using the HMC algorithm,
including those with $\mu=0$, {\it i.e.} no extra fermions
and pseudo-fermions.
The step size $\delta\tau$ of the molecular dynamics
evolution with the leapfrog scheme is 0.01 for all lattices
except for one choice of the action ($S_{Pl}$ with
$1/\epsilon=2/3$ and $\mu=0.4$).
The unit length of the HMC trajectory is set to 0.5.
The total number of the HMC trajectories and observed
acceptance rate $P_{acc}$ are listed in Table~\ref{tab:param}.
Also listed is the integrated autocorrelation time
$\tau_{int}$ measured for the plaquette variable.

\begin{table}[tbp]
  \centering
  \begin{ruledtabular}
  \begin{tabular}{cccccccccc}
    action & $1/\epsilon$ & $\mu$ & $\beta$ & 
    $\delta\tau$ & \#Trj & $P_{acc}$ & $\tau_{int}$ &
    $\langle e^{-\Delta H}\rangle$ & $r_0/a$ \\
    \hline
    Pl & 0   & 0   & 5.83 & 0.01 & 20,000 & 0.815 & 21.6 &
    1.005(4) & 4.27(29)\\
    Pl & 0   & 0.2 & 5.70 & 0.01 & 11,600 & 0.812 &  9.6 &
    0.996(5) & 4.08(10)\\
    Pl & 0   & 0.4 & 5.45 & 0.01 & 11,600 & 0.826 &  6.4 &
    1.000(6) & 3.81(8)\\
    \hline                                                     
    Pl & 2/3 & 0   & 2.33 & 0.01   & 20,000 & 0.849 &  3.6 &
    1.002(3) & 3.84(5)\\
    Pl & 2/3 & 0.2 & 2.23 & 0.01   & 20,000 & 0.844 &  2.1 &
    0.995(3) & 3.94(7) \\
    Pl & 2/3 & 0.4 & 2.06 & 0.0067 & 14,800 & 0.933 &  2.0 &
    1.000(1) & 3.91(11)\\
    \hline
    RG & --- & 0   & 2.43 & 0.01   & 20,000 & 0.781 &  5.1 &
    0.999(4) & 3.91(5)\\
    RG & --- & 0.2 & 2.37 & 0.01   & 21,600 & 0.786 &  4.4 &
    0.998(4) & 3.90(7)\\
    RG & --- & 0.4 & 2.27 & 0.01   & 20,000 & 0.793 &  3.5 & 
    0.999(4) & 3.84(4)\\
  \end{tabular}
  \end{ruledtabular}
  \caption{Simulation parameters.}
  \label{tab:param}
\end{table}

\begin{figure}[tbp]
  \centering
  \includegraphics[width=7cm,clip=true]{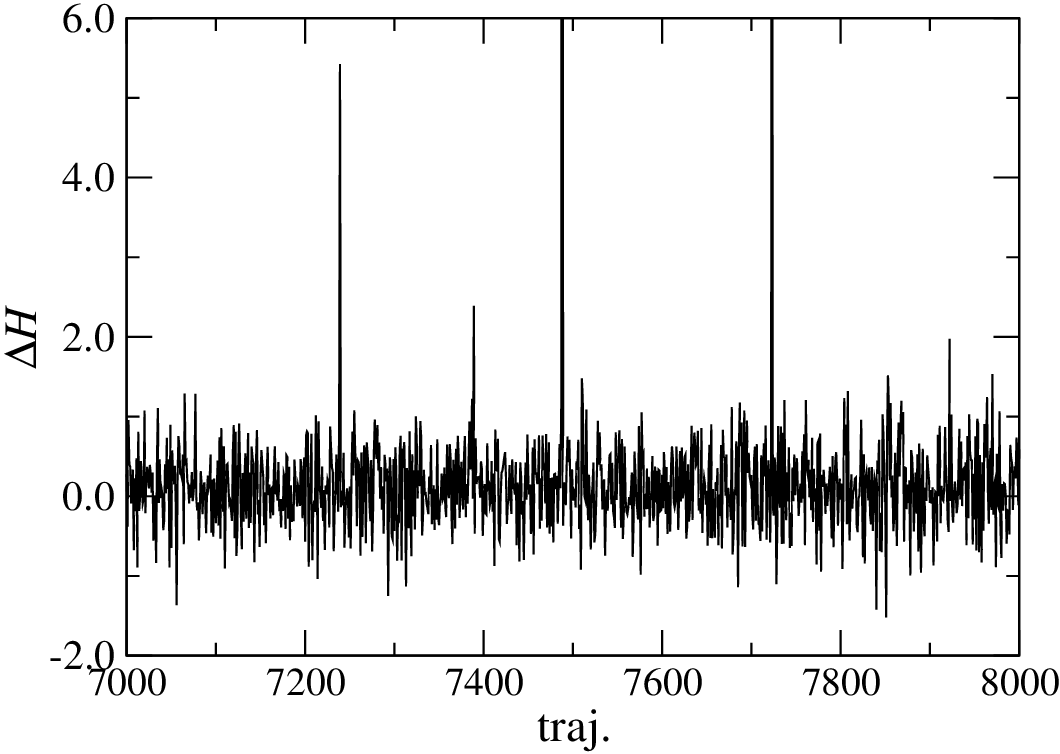}
  \hspace*{2mm}
  \includegraphics[width=7cm,clip=true]{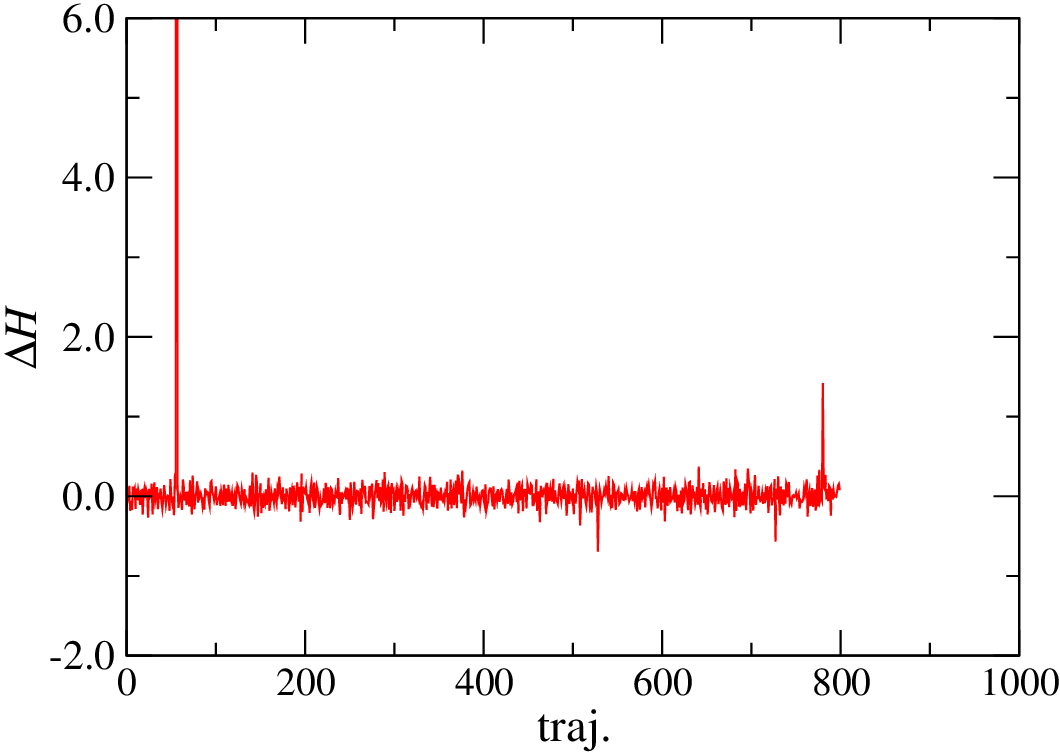}
  \caption{
    History of $\Delta H$ for the lattice 
    ``Pl'' with $1/\epsilon=0$.
    The left panel shows our main run with
    $\delta\tau=0.01$, while the right panel is for a
    reduced step size: $\delta\tau=0.005$.
  }
  \label{fig:deltaH}
\end{figure}

It is known that the HMC simulation becomes unstable due to
(near-)zero eigenmodes
(see, for example, \cite{Namekawa:2004bi}).
Since the fermion force $F$ in molecular dynamics steps
contains a piece that is proportional to $1/\lambda_{min}$, 
the step size $\delta\tau$ must be kept small such that 
the combination $F\times\delta\tau$ is less than $O(1)$.
Because there is no explicit lower bound on $\lambda_{min}$
for the Wilson-type fermions, very small eigenvalue could
appear, though it is suppressed by the fermion determinant.
The problem appears as exceptionally large values of 
$\Delta H$, the difference of the Hamiltonian between the
initial and final steps of the HMC trajectory.
In Figure~\ref{fig:deltaH} we show a typical HMC history of
$\Delta H$ for the lattice ``Pl'' with $1/\epsilon=0$
(standard Wilson gauge action).
On the left panel we show our main run listed in
Table~\ref{tab:param}.
$\Delta H$ is of order one for most trajectories but
develops several large spikes during long runs.
We also made a test run with a reduced step size by a factor
of 2 ($\delta\tau=0.005$) and show the result on the right
panel, for which the fluctuation of $\Delta H$ is reduced
and the occurence of the spikes becomes rare.

The volume dependence of the computational cost for the
extra Wilson fermion is quite normal.
We repeated the test run at $\delta\tau=0.005$ on a smaller
lattice $12^3\times 24$, and measured the average value of
$\Delta H$.
With 1,000 HMC trajectories we obtain 
$\langle\Delta H\rangle$ = 0.0038(23), which may be compared
with the result on a larger lattice 0.0061(59).
Although the statistical error is large, the volume scaling
is consistent with the conventional $O(V)$
\cite{Gupta:1988js} (or milder),
which implies that the total simulation cost scales as
$V^{5/4}$.
There is no significant spike on both lattice at this small
step size.

The exceptional trajectories result in a violation of area
preserving property of the leapfrog integration of the
molecular dynamics evolution.
If it is violated, the detailed balance condition of the
Monte Carlo algorithm cannot be proved, and therefore the
exactness of the algorithm is lost.
This is a problem associated with a discontinuity in the
effective pseudo-fermion action, as in the case for
(\ref{eq:pseudo-fermion}).
If the step size is small enough, the trajectory would never
pass the discontinuity because of the strong repulsive force
from the potential wall, and no problem arises.
But for some large step size the trajectory may occasionally
go across the discontinuity, and then the conservation of
the Hamiltonian is strongly violated.

This problem can be monitored by the quantity
$\langle e^{-\Delta H}\rangle$, which must be consistent
with 1 when the area preserving property is satisfied.
In the second last column of Table~\ref{tab:param} we list
the value of $\langle e^{-\Delta H}\rangle$.
For most cases it is consistent with 1 within one standard
deviation.
The problem is expected to disappear if one chooses small
enough step size $\delta\tau$.
In fact, for the run ``Pl'' with $1/\epsilon=2/3$ and
$\mu=0.4$, for which $\delta\tau$ is 2/3 of other cases, the
number of large spikes is much reduced and the relation
$\langle e^{-\Delta H}\rangle=1$ is satisfied to a good
precision. 

In the following analysis we assume that the gauge
configurations are properly sampled.
Even if the exceptional trajectories exist, they are almost
always rejected by the Metropolis test and thus do not
affect the following trajectories.
In the productive run with the dynamical overlap fermion,
that we are currently carrying out, the step size is 
carefully chosen to avoid the potential problem.

\section{Spectral density}
\label{sec:spectral_density}
We investigate the effect of the extra Wilson fermions on
the spectral density of the hermitian Wilson-Dirac operator
$H_W(m_0)$ in the quenched approximation.

We take 30--100 gauge configurations for each runs listed in
Table~\ref{tab:param} and calculate the low-lying
eigenvalues of $H_W(m_0)$ with $m_0=-1.6$.
For each runs the gauge coupling is chosen such that the
lattice spacing is roughly tuned to 0.125~fm.
Statistical correlation between consecutive gauge
configurations is negligible, as they are separated by 200
HMC trajectories.
We use the conventional Lanczos algorithm, and calculate the 
eigenvalues with its absolute value less than 0.2 in the
lattice unit.

\begin{figure}[tbp]
  \centering
  \includegraphics[width=8cm,clip=true]{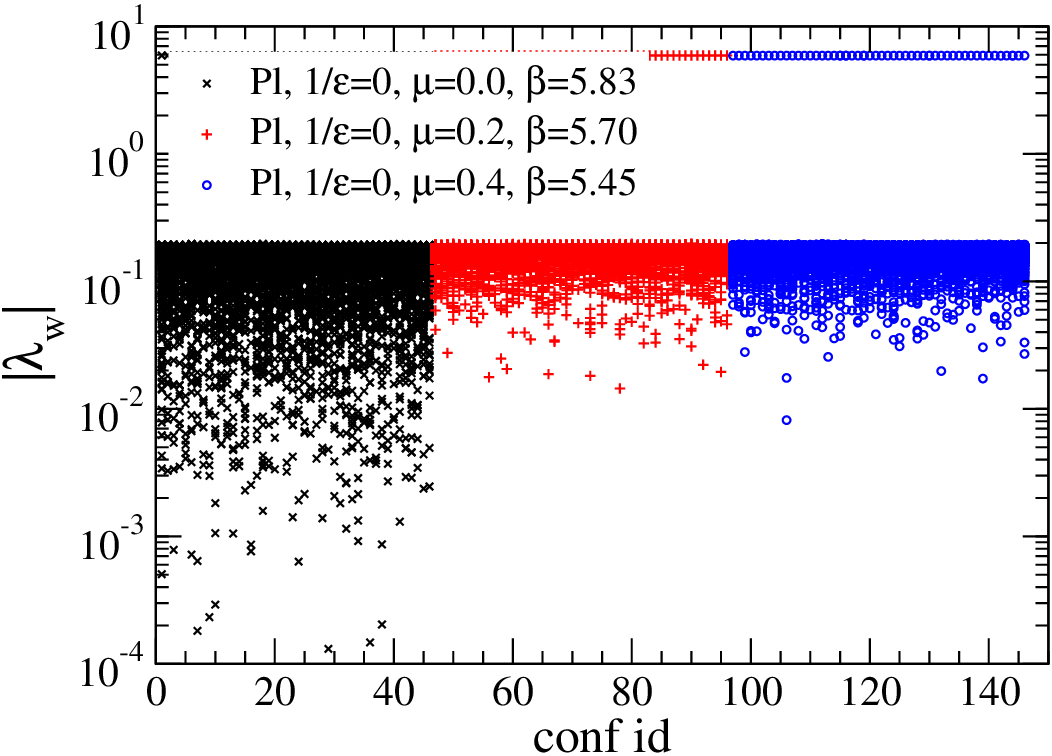}\\[2mm]
  \includegraphics[width=8cm,clip=true]{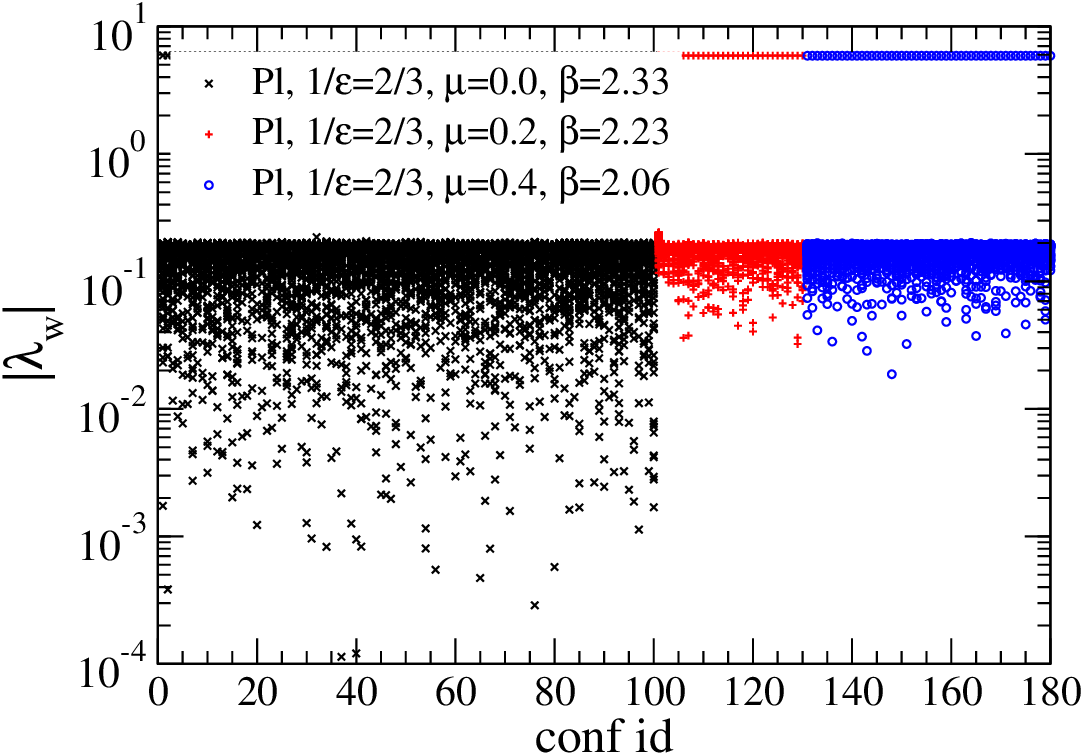}\\[2mm]
  \includegraphics[width=8cm,clip=true]{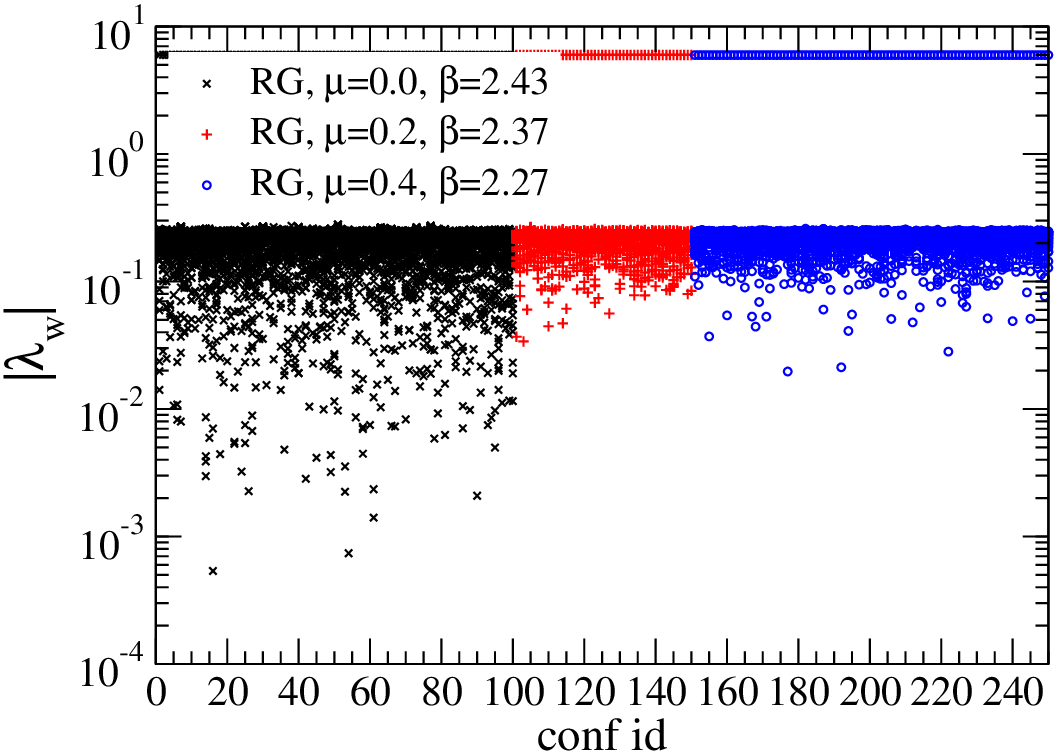}
  \caption{
    Low-lying eigenvalues of $H_W(m_0)$.
    Lattice actions are
    ``Pl'' with $1/\epsilon=0$ (top),
    ``Pl'' with $1/\epsilon=2/3$ (middle), and
    ``RG'' (bottom).
    Data for three values of $\mu$ ($\mu$ = 0.0, 0.2, and
    0.4) are shown in each plot.
    The highest mode is also shown, whose value is $\sim$
    5.9, almost independent of the gauge configuration.
    }
  \label{fig:eigen_dist}
\end{figure}

In Figure~\ref{fig:eigen_dist} we plot the absolute value of
the observed eigenvalues $\lambda_W$ for each set of
configurations.
Different choices of lattice actions are shown in separate
panels, and in each panel the results at three values of
$\mu$ (0.0, 0.2, and 0.4) are plotted.
As one can clearly see, there are significant number of
eigenvalues $|\lambda_W|$ less than 0.01 for any of the
three gauge actions if the extra Wilson fermions are not
included ($\mu$ = 0).
On the other hand, with the finite values of $\mu$, the
eigenvalues less than 0.02 do not appear at all.
The difference between $\mu$=0.2 and 0.4 is not significant,
but $\mu$=0.2 seems slightly better.

\begin{figure}[tbp]
  \centering
  \includegraphics[width=8cm,clip=true]{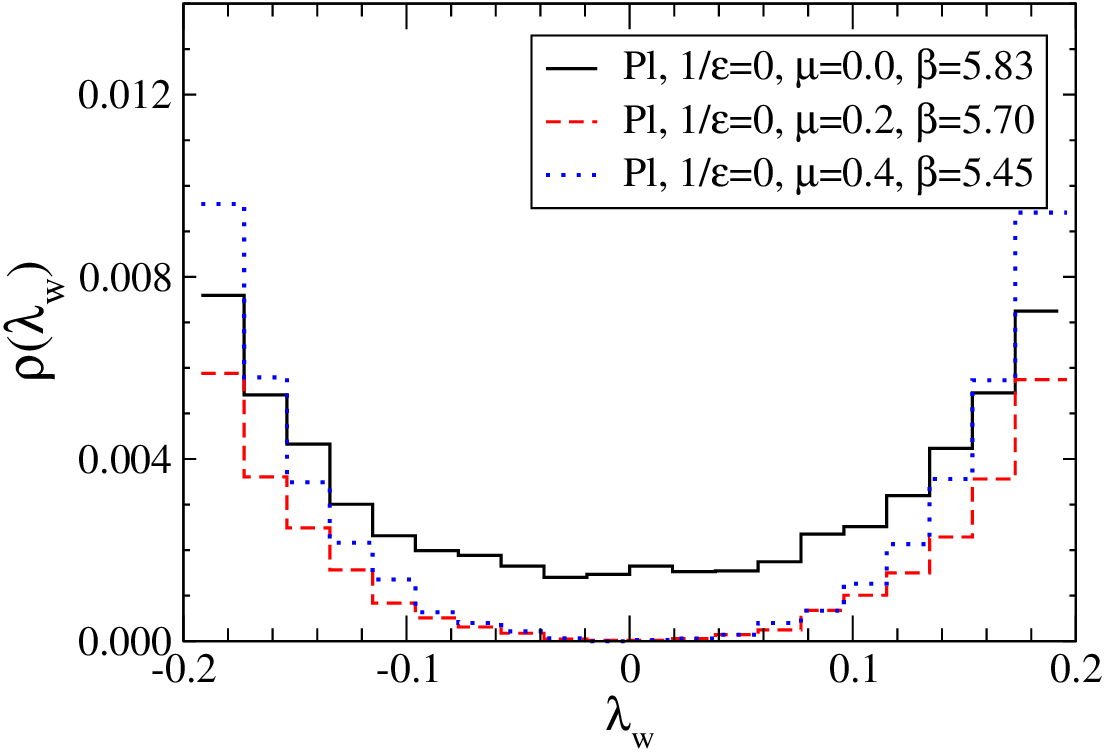}\\[2mm]
  \includegraphics[width=8cm,clip=true]{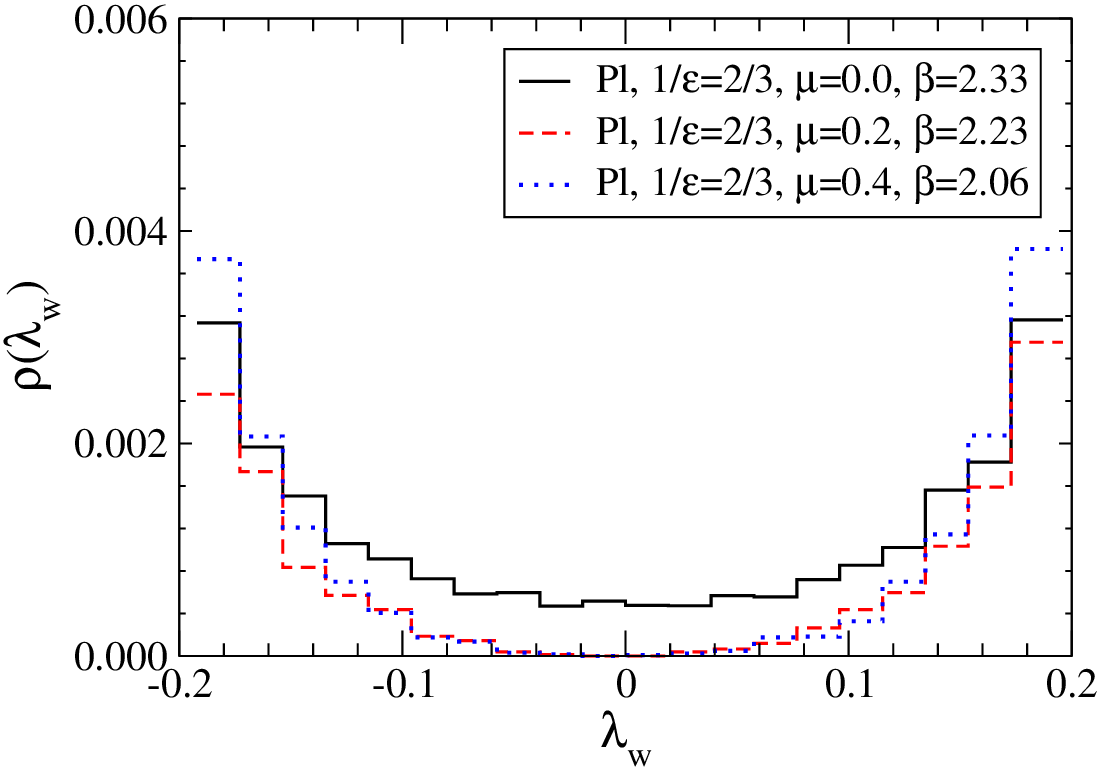}\\[2mm]
  \includegraphics[width=8cm,clip=true]{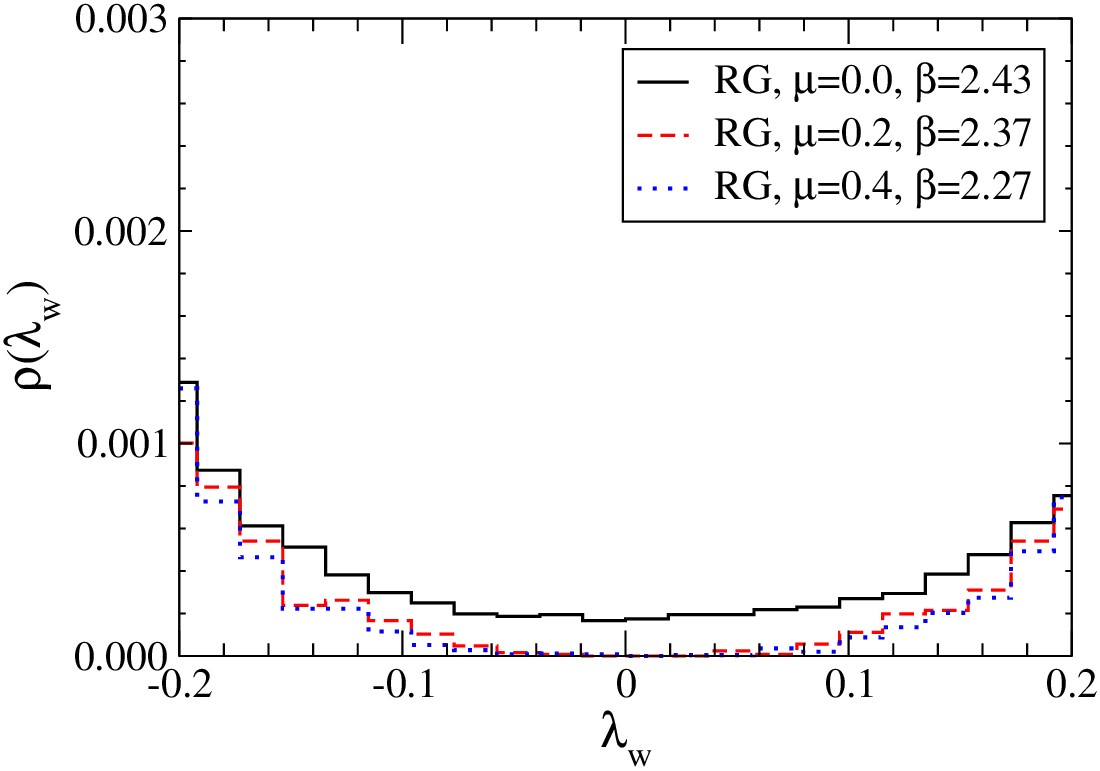}
  \caption{
    Histogram of the spectral density of $H_W(m_0)$.
    Lattice actions are
    ``Pl'' with $1/\epsilon=0$ (top),
    ``Pl'' with $1/\epsilon=2/3$ (middle), and
    ``RG'' (bottom).
    Data for three values of $\mu$ ($\mu$ = 0.0, 0.2, and
    0.4) are shown in each plot.
    }
  \label{fig:eigen_hist}
\end{figure}

These observations can be made more quantitative using the
spectral density $\rho(\lambda_W)$.
The spectral density is defined by 
\begin{equation}
  \label{eq:spectral_density}
  \rho(\lambda_W) = \frac{1}{V}
  \left\langle
    \sum_n \delta(\lambda_W-\lambda_n)
  \right\rangle,
\end{equation}
where $\langle\cdots\rangle$ denotes the ensemble average
and $\lambda_n$ represents each eigenvalue on a given gauge
configuration.
In Figure~\ref{fig:eigen_hist} we plot the histogram of
$\rho(\lambda_W)$.
Each bin in the plots has a size of 0.0192 in $\lambda_W$.
We observe non-zero spectral density at $\lambda_W=0$ when
there is no extra fermion introduced ($\mu=0.0$).
We calculate the value of $\rho(0)$ by fitting an integrated
density
$I(\bar{\lambda})=\int_{-\bar{\lambda}}^{+\bar{\lambda}}
 d\lambda \rho(\lambda)$
with a polynomial
$2\rho(0)\bar{\lambda}+O(\bar{\lambda}^3)$.
The results are
$\rho(0)=1.509(3)\times 10^{-3}$ for ``Pl'' with $1/\epsilon=0$,
$0.494(1)\times 10^{-3}$ for ``Pl'' with $1/\epsilon=2/3$, and
$0.177(1)\times 10^{-3}$ for ``RG''.
As well known, the density of the near-zero modes is
substantially reduced for the improved actions.

With the extra Wilson fermions the near-zero mode density is
essentially zero; there is no events in the lowest bin.
The curve of $\rho(\lambda_W)$ near $\lambda_W=0$ is
consistent with $\sim\lambda_W^2$ for both $\mu=0.2$ and
0.4.
The difference between $\mu=0.2$ and 0.4 is not significant,
except for the highest bin at $|\lambda_W|\simeq 0.2$.

Among the three choices of the gauge actions, the
suppression of the near-zero modes is most efficient for the
RG action.
We should note that the scale of the vertical axis in
Figure~\ref{fig:eigen_hist} is different among the three
plots, and the value at $|\lambda_W|\simeq 0.2$ is 6--10
times smaller for the RG action than the standard Wilson
gauge action ($1/\epsilon=0$).
The choice ``Pl'' with $1/\epsilon=2/3$ is in between the
other two.
Therefore, we may conclude that the rectangular term is more
effective to suppress the dislocation than restricting the
range of the plaquette variable by the denominator in
(\ref{eq:admiaction}). 

\section{Topology conservation}
\label{sec:topology_conservation}
As we discussed in the Introduction, the net topological
charge defined as an index of the overlap-Dirac operator
cannot change by the continuous deformation of the gauge
field variables when the extra Wilson fermion is included. 
This is approximately the case in our simulation using the
HMC algorithm with small molecular dynamics time step
$\delta\tau$.
The large spikes in $\Delta H$ as discussed in
Section~\ref{sec:HMC_simulations} may indicate some attempts
of the gauge field to go beyond the potential barrier, but
the acceptance probability of such trajectories is
essentially zero and therefore the topology does not change
in the accepted trajectories.

In order to explicitly check the topological charge of the
generated gauge configurations, we calculate the near-zero
eigenmodes of the overlap-Dirac operator $D$ as defined in
(\ref{eq:ov}).
The kernel operator is the Wilson-Dirac operator with
$m_0=-1.6$, the same value as in the extra Wilson fermion.
The topological charge can be identified as the number of
the left-handed or right-handed zero modes.
In practice we adopt the method proposed in
\cite{Giusti:2002sm}.
Namely, we calculate the low-lying eigenvalues of chirally
projected operators $D^\pm\equiv P_\pm DP_\pm$.
Since the non-zero eigenvalues appear in pair between $D^+$
and $D^-$, the remaining unpaired eigenvalues very close to
zero can be identified as chiral zero modes.

We calculate the eigenvalues of $D$ on 80 configurations
of ``RG'' with $\mu=0.2$, and find no exact chiral zero
mode.
This is consistent with our expectation, as these
configurations are generated starting from an initial
configuration in the trivial topological sector, {\it i.e.}
$Q=0$.
In addition, we also generated another set of
configurations starting from a $Q=-2$ configuration (number
of samples is 60).
On all of these configurations we confirm the presence of
two left-handed zero modes as expected.
From these observations we conclude that the topological
charge is indeed conserved during the HMC simulations.
We are currently performing many set of simulations
including dynamical overlap fermions as well as the extra
Wilson fermions. 
Also in these simulations, there has been no sign of the
change of the topological charge so far.

\section{Topology dependence}
\label{sec:topology_dependence}
Since the topological charge conserves during the HMC
simulation, one cannot sample the correct $\theta=0$ vacuum,
for which the topological charge distributes according to
the topological susceptibility.
However, the cluster decomposition property of the local
field theory suggests that the physical quantities measured
in a local space-time region do not depend on the
topological fluctuations occuring far apart from that region.
This means that the physical quantities do not depend on the
global topological charge, as far as the space-time volume
is large enough.
The topological fluctuation is controlled
by the topological susceptibility 
$\chi_t\equiv\langle Q^2\rangle/V$, which is proportional to
$m\Sigma=M_\pi^2 F_\pi^2$ in unquenched QCD near the chiral
limit. 
The typical length scale for the topological fluctuation is
then given by $(M_\pi F_\pi)^{-1/2}$.
The lightest possible pion mass in our planned dynamical
simulation is $M_\pi\simeq$ 300~MeV, for which
the typical scale is $\sim$ 1~fm.
It becomes even smaller for heavier sea quark masses or for
the quenched theory. 
Therefore, our naive expectation is that the fixed-topology
simulations in a $(2\mbox{~fm})^4$ box or larger do not have
too much impact on physical quantities.

On a rather general ground, one can prove that the hadron
masses in a fixed topological sector deviate from the
correct value in the $\theta=0$ vacuum by 
$O(1/\langle Q^2\rangle)$, which scales as $\sim 1/V$
\cite{Brower:2003yx}.
An estimate of the coefficient using the chiral effective
lagrangian implies that the deviation of pion mass does not
exceed 1\% even for light pions ($\sim$ 300~MeV)
on a $(2\mbox{~fm})^4$ box, and the effects on other hadrons
are even smaller.

In this paper we show a study in the quenched approximation,
instead of the time-consuming dynamical fermion simulations.
In the quenched theory, however, the situation could be
drastically different because the hadron masses are more
infra-red sensitive due to the double pion pole structure at
the one-loop level of quenched chiral perturbation theory.
Because of the integral of the form
$\int\! d^4p\, m_0^2/(p^2+m_\pi^2)^2$, the diagram is infra-red
divergent in the chiral limit, and the information of the
topological fluctuation of the whole space-time is gathered 
through the singlet vertex $m_0^2$ proportional to the
topological susceptibility $\chi_t$, whose integral over
space-time gives $Q^2$ the fixed topological charge in this
case. 
We therefore expect the quenched chiral logarithm that
depends on the topological charge as $Q^2/V$ rather than its
average $\langle Q^2\rangle/V$.

\begin{figure}[tbp]
  \centering
  \includegraphics[width=8cm,clip=true]{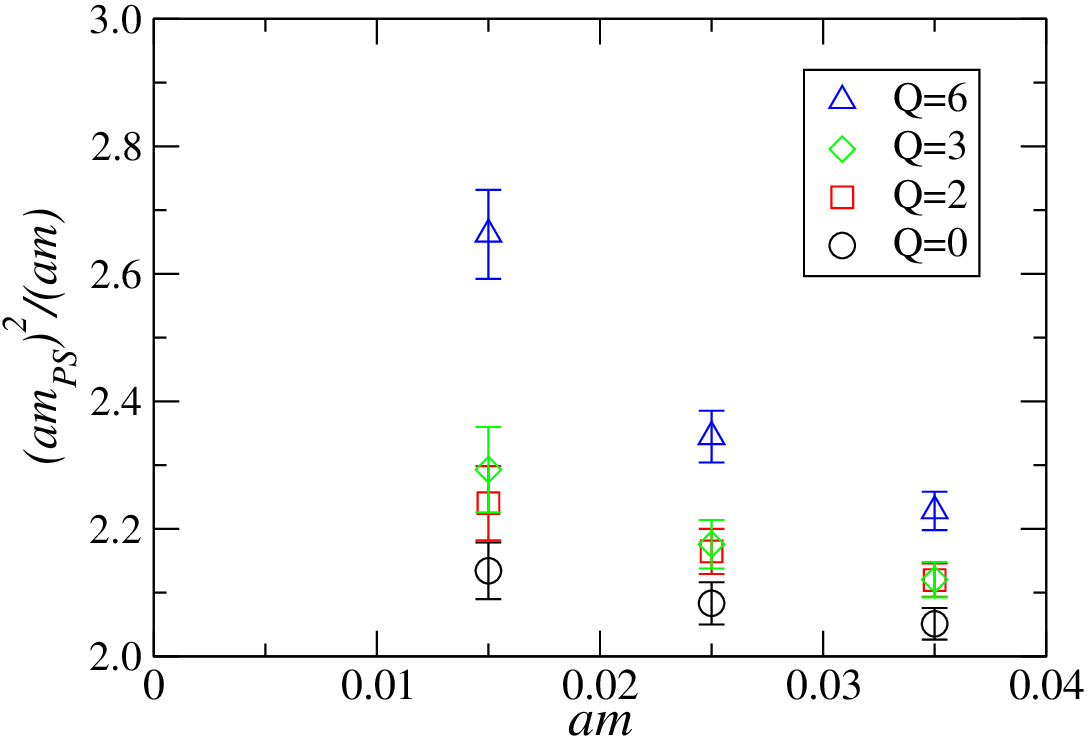}
  \caption{
    Pseudo-scalar mass squared calculated on the
    quenched configurations at fixed topological charges
    $|Q|$ = 0, 2, 3, and 6.
  }
  \label{fig:mPS2}
\end{figure}

In order to see the $Q$ dependence in the simulations, we
carry out a calculation of pion mass on quenched gauge
configurations with fixed topological charges $|Q|$ = 0, 2,
3, and 6.
The gauge configurations are generated with the extra Wilson
fermions at $\beta=2.37$ (RG gauge action), and the
statistics is 100 for each topological charge.
The results for $(am_{PS})^2/(am)$ are plotted in
Figure~\ref{fig:mPS2} as a function of the quark mass $am$.
We find rather large topological charge dependence as
expected in the quenched theory:
the difference of $am_{PS}$ between $|Q|=0$ and 6 is as
large as 10\%.
A similar dependence is also found in \cite{Galletly:2006hq}.

Such a large topology dependence is not expected for
unquenched QCD, because there is no pathological infrared
divergence. 
In our on-going dynamical overlap fermion project, we are
planning to investigate the topology dependence in detail as
well as the volume dependence.

\section{Beta shift}
\label{sec:beta_shift}
The extra Wilson fermions and their associated
pseudo-fermions have masses of order of the lattice cutoff
and do not affect the physics at low energy.
Their only effects are ultraviolet ones, such as the
renormalization of the strong coupling constant and quark
masses.
Here we calculate the finite renormalization of the strong
coupling constant (or the $\beta$ shift) due to the extra
fields at the one-loop level in perturbation theory.
Such calculation for the massless Wilson fermion was done
before \cite{Hasenfratz:1993az,Lee:1998ng}, but we must
repeat the calculation with the large negative mass.

The shift of the coupling constant is given as
\begin{equation}
  \frac{1}{g_{\mathrm{eff}}^{(N_f)2}} = 
  \frac{1}{g_0^2} - \Pi(p),
\end{equation}
where $\Pi(p)$ is the vacuum polarization function of gluon
due to fermions:
\begin{equation}
  \Pi_{\mu}(p) = (p^2\delta_{\mu\nu}-p_\mu p_\nu)\Pi(p).
\end{equation}
For the usual (lattice) fermions, it is written as
\begin{equation}
  \label{eq:Pi}
  \Pi(p) = N_f 
  \left[ \frac{1}{24\pi^2}\ln(a^2p^2) + k_f \right], 
\end{equation}
and the constant $k_f$ depends on the fermion formulations.
One-loop calculation gives
$-$0.013732 (Wilson fermion) \cite{Weisz:1980pu,Kawai:1980ja},
$-$0.038529 (clover fermion) \cite{Sint:1995ch}.
The logarithmic term in (\ref{eq:Pi}) corresponds to the
usual one-loop running of the coupling constant and gives
the $N_f$ dependent coefficient of the beta function
$b_0=\frac{1}{(4\pi)^2}\left(\frac{11}{3}N-\frac{2}{3}N_f\right)$.

\begin{figure}[tbp]
  \centering
  \includegraphics[width=8cm,clip=true]{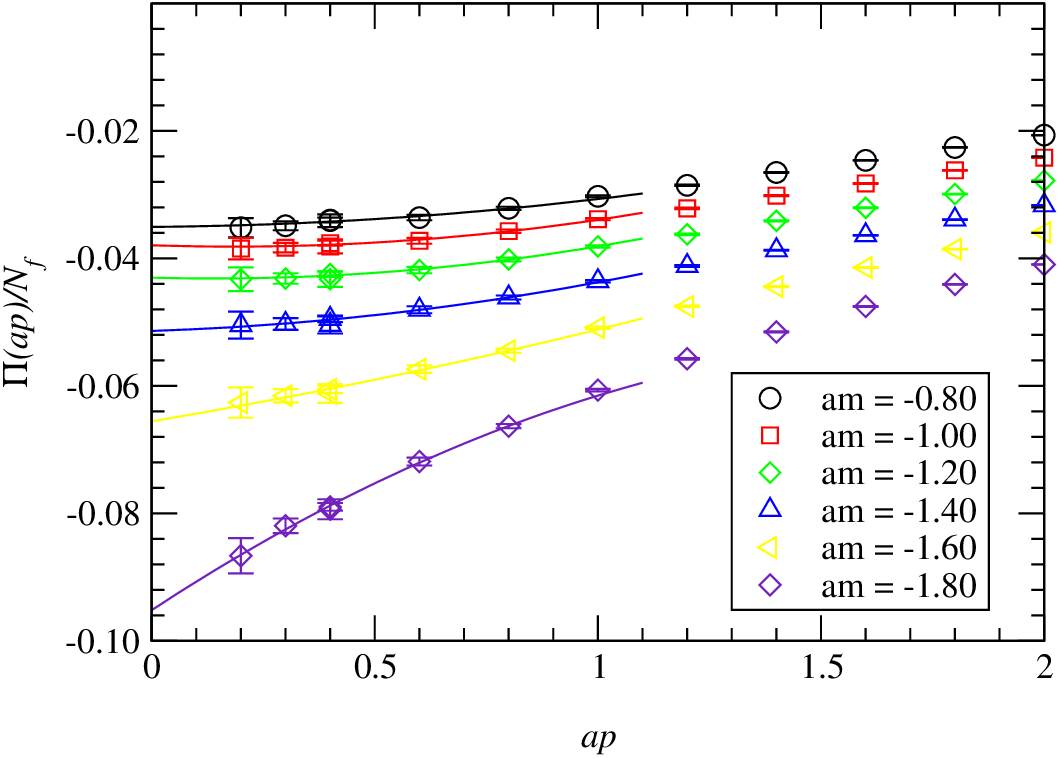}
  \caption{$\Pi(p)/N_f$ for various mass parameters.}
  \label{fig:Pi}
\end{figure}

The unphysical fermions with a large negative mass 
do not contribute to the logarithmic term in (\ref{eq:Pi}).
Therefore, the $p\to 0$ limit is finite.
Figure~\ref{fig:Pi} shows $\Pi(p)/N_f$ for the negative mass
Wilson fermion with masses $am=-0.80\sim-1.80$.
Since the numerical integral becomes unstable near $ap=0$,
we obtain the $ap\to 0$ limit by an extrapolation using the
data points below $ap=1$.
The numerical results for $\Pi(0)$ are listed in
Table~\ref{tab:Pi_nm}. 

At $am_0=-1.6$ the $\beta$ shift $\delta\beta=-6\Pi(0)$
is calculated as 0.786(4) at the one-loop level.
It means that we have to reduce the $\beta$ value by this
amount in order to simulate without changing the lattice
spacing. 
The actual value could be significantly larger due to higher 
order corrections.
With the mean field improvement, for instance, the value is
divided by the expectation value of the plaquette 
$\langle P\rangle$, which is typically around 0.6 for the
Wilson gauge action at $\beta=6$.

\begin{figure}[tbp]
  \centering
  \includegraphics[width=8cm,clip=true]{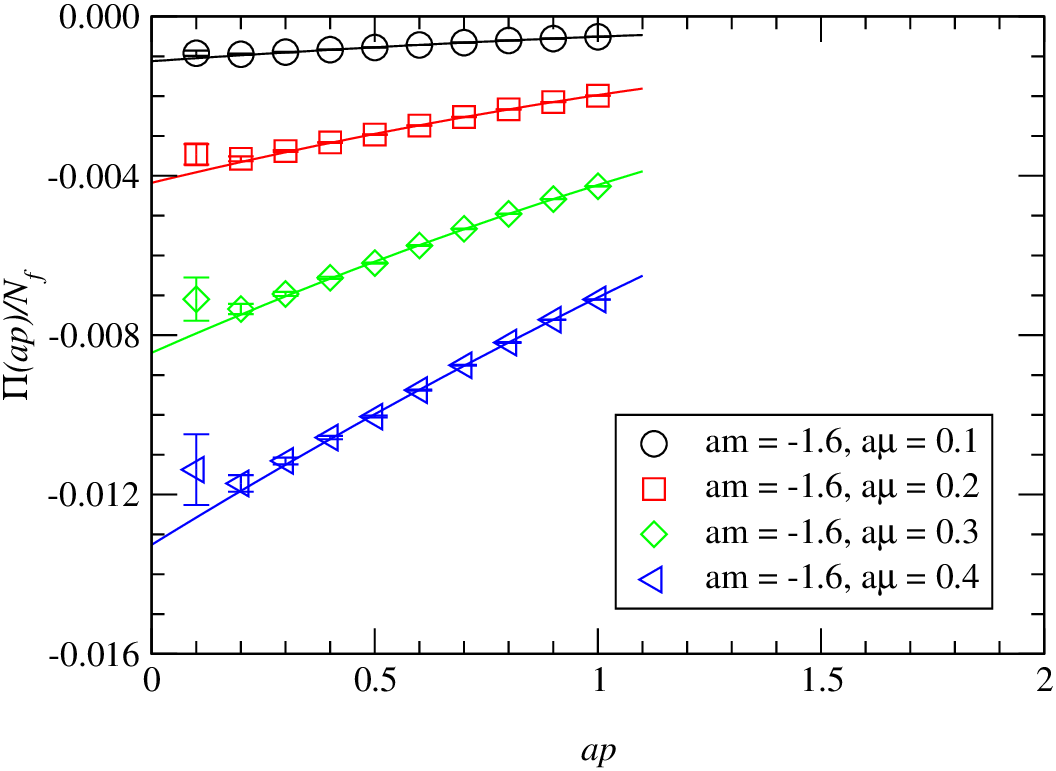}
  \caption{$\Pi(p)/N_f$ including the twisted mass pseudo-fermions.}
  \label{fig:tm}
\end{figure}

\begin{table}[tbp]
  \centering
  \begin{ruledtabular}
  \begin{tabular}{cccccc}
    $am_0$ & w/o pseudo-fermion & 
    $\mu=0.1$ &
    $\mu=0.2$ &
    $\mu=0.3$ &
    $\mu=0.4$\\
    \hline
    $-$0.8 & $-$0.0351(2)\\
    $-$1.0 & $-$0.0379(2) &
    $-$0.000395(3) & $-$0.00153(1) &
    $-$0.00329(3) & $-$0.00549(5) \\
    $-$1.2 & $-$0.0430(3) &
                  & $-$0.00171(1) \\
    $-$1.4 & $-$0.0514(3) &
                  & $-$0.00228(2) \\
    $-$1.6 & $-$0.0655(3) & 
    $-$0.00112(1) & $-$0.00418(3) &
    $-$0.00844(7) & $-$0.0133(1) \\
    $-$1.8 & $-$0.0952(4) &
                  & $-$0.01209(4) 
  \end{tabular}
  \end{ruledtabular}
  \caption{Vacuum polarization function $\Pi(p)/N_f$ at $p=0$.}
  \label{tab:Pi_nm}
\end{table}

Including the twisted mass pseudo-fermions, the $\beta$
shift is substantially reduced.
It vanishes in the limit of $\mu\to 0$ and the 
contribution for finite $\mu$ starts from $\mu^2$.
Results are shown in Figure~\ref{fig:tm} for $am_0=-1.6$ and
the twisted mass $\mu$ = 0.1--0.4.
Note that the vertical axis is one order of magnitude
smaller than that in Figure~\ref{fig:Pi}.
At $\mu$ = 0.2, the $\beta$ shift is 0.0502(4) ($N_f$ = 2),
which may be compared with the actual data:
$\beta=5.83\to 5.70$ for ``Pl'' with $1/\epsilon$=0,
$2.33\to 2.23$ for ``Pl'' with $1/\epsilon$=2/3, and
$2.43\to 2.37$ for ``RG''.
The one-loop calculation underestimates the measured value
by about a factor of two for the plaquette gauge action,
while it gives a good approximation for the RG action.
This is consistent with the expected mean field enhancement
of the coupling constant for the plaquette gauge action.

The small $\beta$ shift for the cases with the twisted mass
pseudo-fermions is desirable, because one can avoid too small
$\beta$ values in the dynamical fermion simulations.
In particular, for the Wilson gauge action there is a
remnant of the fundamental-adjoint phase transition
\cite{Bhanot:1981pj} in the strong coupling regime
($\beta\lesssim 5.2$).
With dynamical fermions it may appear as a real first-order
phase transition, which prevents one from taking smooth
continuum limit.
Examples are found in the simulations with two
\cite{Blum:1994eh,Farchioni:2004us} 
and three flavors \cite{Aoki:2004iq} of the Wilson-type
fermions.

\section{Concluding remarks}
\label{sec:conclusions}
By introducing the extra Wilson fermions with the large
negative mass, it is possible to provide a gap in the
spectral density of $H_W$ and thus to remove the problem
associated with the near-zero modes.
For the overlap fermion the exponential localization of the
Dirac operator is guaranteed; for the domain-wall fermion
the residual chiral symmetry breaking will be much reduced.
For three choices of gauge actions, we confirm that the
spectral density of $H_W$ vanishes at $\lambda_W=0$ and the
occurrence of the near-zero mode is very much suppressed.
The number of remaining near-zero modes is smallest for the
Iwasaki gauge action.
By also adding the twisted mass pseudo-fermions, the effect
on the coupling renormalization can be minimized while
keeping the good property of suppressing the near-zero
modes.

A remaining problem concerning the locality of the
overlap-Dirac operator is its actual localization length in
the numerical simulations.
For the numerical simulation with controlled discretization
error, the localization length has to be much smaller than
$1/\Lambda_{\mathrm{QCD}}$, and 
this gives a stronger practical upper limit of the gauge
coupling in addition to the more fundamental limit from
locality. 
Golterman, Shamir, and Svetitsky argued that the
localization length is controled by the mobility edge of the
eigenmodes of $H_W$
\cite{Golterman:2003qe,Golterman:2004cy,Golterman:2005fe}.
We are currently calculating the spatial correlation on the
low-lying mode eigenvectors in order to determine the
mobility edge. 
This work will be published elsewhere.

Although the extra Wilson fermion is also useful in the
quenched QCD simulations, our main objective is to perform
the unquenched simulations using the fermion formulation
with exact chiral symmetry.
So far, the dynamical overlap fermion simulation has been
attempted only on small lattices
\cite{Fodor:2003bh,DeGrand:2004nq,DeGrand:2005vb,Cundy:2005pi}, 
as its computational cost is extremely large.
One of the reasons for the large numerical cost is the
treatment of the sign function.
When the low-lying eigenvalue $H_W$ passes zero during the
molecular dynamics evolutions, the sign function changes its
value discontinuously and its derivative diverges.
Fodor, Katz and Szabo \cite{Fodor:2003bh} introduced
so-called the reflection/refraction process on the
$\lambda_W=0$ surface, which requires monitoring of the
low-lying eigenvalues and takes lots of computational costs.
The cost would scale as $V^2$, as the lattice volume $V$ is
increased, and therefore the simulation on large lattices
could be prohibitively costly.
This problem can be totally avoided by the extra Wilson
fermion, because the zero crossing never happens if the
molecular dynamics step size is chosen small enough.
The extra cost for the Wilson fermion is negligible in 
the dynamical overlap fermion simulation.
Using this lattice action, we are currently carrying out the
dynamical overlap fermion simulations on a $16^3\times 32$
lattice.

An immediate question on the strategy of suppressing the
low-lying modes of $H_W$ is that the simulation is trapped
in a given topological sector and one cannot achieve the
correct sampling of the $\theta=0$ vacuum of QCD.
This is an algorithmic problem of the molecular dynamics
simulation, in which the gauge variables are changed
continuously along trajectories.
We first emphasize that the continuum theory has the same
property, {\it i.e.} the topological nature of the gauge
field on a torus.
Any lattice gauge action should recover this topological
property as the continuum limit is approached, and our 
choice has this property at any finite lattice spacing.
The configuration space of a given fixed topology is simply 
connected in the continuum limit, and therefore the
ergodicity of the Monte Carlo simulation is satisfied even
with a fixed topology.
At finite lattice spacings there is no proof of the
ergodicity, but there is no indication of its violation
either.
In order to detect such an effect if any, we should probably
look at quantities sensitive to the topological charge
density. 
That is an interesting subject, which we leave for future
studies. 

Second, because of the cluster decomposition principle any
physical quantity should not depend on the net topological
charge, if the lattice volume is large enough.
Even on a lattice with a fixed topological charge, local
topological fluctuation could occur at different positions
on the lattice keeping the net topological charge constant, 
and the frequency of the topological fluctuation is
characterized by the topological susceptibility.
In other words, the effect of fixing the net topology is a
finite volume effect.
(An important exception to this statement is the CP-odd
quantities, such as the neutron electric dipole moment.)
As we discussed in Section~\ref{sec:topology_dependence},
hadron masses calculated at a fixed topology
contains a finite volume correction of $O(1/V)$ to the
physical value in the $\theta=0$ vacuum
\cite{Brower:2003yx}. 
Although the effect is quite large in the quenched QCD due
to the sickness of quenched theory, we expect a small effect
on dynamical lattices.
Study of the topological charge dependence is underway as a
part of our project of dynamical overlap fermion
simulations.

\begin{acknowledgments}
  Numerical simulations are peformed on Hitachi SR11000 and
  IBM BlueGene/L at High Energy Accelerator Research
  Organization (KEK) under a support of its
  Large Scale Simulation Program No.~3 (FY2006).
  At an early stage, we also used NEC SX8 at Yukawa
  Institute for Theoretical Physics (YITP), Kyoto University.
  The authors thank YITP, where this work was initiated
  during the YITP workshop on ``Actions and symmetries in
  lattice gauge theory'' (YITP-W-05-25).
  H.~F. and T.~O. thank M.~L\"uscher for useful discussions.
  N.~Y. thank T.~Blum for discussions.
  This work is supported in part by the Grant-in-Aid of the
  Ministry of Education 
  (No.~1315213, No.~16740147, No.~16740156,
  No.~17740171, No.~18340075, No.~18034011, and No.~18740167).
\end{acknowledgments}

\end{document}